\documentclass[12pt]{article}
\usepackage[T1]{fontenc}
\usepackage[utf8]{inputenc}
\newcommand{\R}   {{\mbox{R\hskip-0.9em{}I \ }}}
\newcommand{\N}   {{\mbox{N\hskip-0.9em{}I \ }}}
\newcommand{\C}   {{\mbox{C\hskip-0.5em{}I \ }}}

\usepackage{amsfonts}
\usepackage{amsmath}
\usepackage{amssymb}
\usepackage{mathrsfs}
\usepackage{latexsym}
\usepackage{multicol}
\usepackage{fancybox}
\usepackage{dsfont}
\usepackage{color}
\usepackage{hyperref}
\usepackage{subfigure} 
\usepackage{graphicx} 
\usepackage{caption}
\usepackage[us,12hr]{datetime}                    %
\usepackage{fancyhdr}                                  %  RIGHE AGGIUNTE PER FARE COMPARIRE
\def\be{\begin{equation}}
\def\ee{\end{equation}}

\def\d'{``}

\newtheorem{thm}{Theorem}[section]

\newtheorem{propn}[thm]{Proposition}
\newtheorem{lemma}[thm]{Lemma}

\newtheorem{cor}[thm]{Corollary}

\def\be{\begin{equation}}
\def\ee{\end{equation}}
\def\bea{\begin{eqnarray}}
\def\eea{\end{eqnarray}}

\def\i'{\textrm{i}}

\begin{document}

\begin{center}
\Large{\bf{Ermakov-Pinney and Emden-Fowler equations: \\
new solutions from novel B\"acklund transformations}}
\end{center}

\begin{center}
{ \large{Sandra Carillo\footnote{Dipartimento Scienze di Base e Applicate per l'Ingegneria, 
\textsc{Sapienza}  Universit\`a di Roma, ROME, Italy \&
 I.N.F.N. - Sezione Roma1
Gr. IV - Mathematical Methods in NonLinear Physics.
} and Federico Zullo\footnote{Dipartimento di Ingegneria Meccanica e Aerospaziale, \textsc{Sapienza}  Universit\`a di Roma, ROME, Italy.}  }}

{ }

\end{center}

\medskip
\medskip

\begin{abstract}
\noindent
The class of nonlinear ordinary differential equations $y''y=F(z,y^2)$, where $F$ is a smooth function,  is studied. Various nonlinear ordinary differential equations, whose applicative importance
is well known, belong to such a class of nonlinear ordinary differential equations. Indeed, the Emden-Fowler equation, the Ermakov-Pinney equation and the  generalized Ermakov equations  are among  them.
B\"acklund transformations and auto B\"acklund transformations are constructed: these last transformations induce the construction of  a ladder of new solutions adimitted by the given differential equations starting from a trivial solutions.  
Notably,  the highly nonlinear structure of this class of nonlinear ordinary differential equations implies that 
numerical methods are very difficulty to apply. 
\end{abstract}

\bigskip\bigskip

\noindent

\noindent
{\bf Keywords}: Non-linear ordinary differential equations, B\"acklund transformations, \\
Schwartzian derivative, Ermakov-Pinney equation, Emden-Fowler equation.

\section{Introduction} \label{intro}
Since the 1990s,  a number of  results on discretization of ordinary differential equations describing integrable physical 
systems were achieved. Among them there are the Ruijsenaars-Schneider model \cite{Nijhoff}, the Henon-Heiles, Garnier 
and Neumann systems \cite{HKR}, \cite{Tsi1}, the Ermakov-Pinney equation \cite{AH}, the Euler top \cite{Bob}, the Lagrange 
\cite{KPR} and Kirchoff tops \cite{Rag2} , the Chaplygin ball \cite{Tsi2}, the Gaudin systems \cite{Rag1}, \cite{FZ2} (see also 
\cite{FZ1} and references therein). All these discretizations represent maps among solutions of the differential equations 
describing the corresponding integrable systems. Usually, these transformations are obtained from the Lax representation of 
the model, by constructing an appropriate {\it dressing matrix} intertwining the Lax representation of the system (see e.g. 
\cite{FZ1} for details). These transformations turn out to be canonical in the phase space of the system. Furthermore, they 
enjoy some crucial properties inherited by the Lax representation and by the integrable structure of the system. In 
particular the maps are algebraic, and, if the Lax matrix is of order $N$, this algebraic structure enters
 through an auxiliary variable 
$\gamma$ which satisfies an irreducible polynomial equation of degree $N$ whose coefficients are rational functions of the 
dynamical variables (see e.g. \cite{HRZ}).\\

In this work an algebraic map among solutions of a class of nonlinear second order differential equations is obtained. 
Integrability of the equations is not requested, neither  the transformations are obtained via the Lax representation of the 
system. Rather, both the differential equations considered and the transformations obtained are an extension of the results given
in \cite{JMAA2000}: in that work the properties of the Schwarzian derivative were fundamental to 
achieve the results. In  the present case, the Schwarzian derivative  continues to  play a key role in the 
construction of the transformations, allowing us to derive new results.

The present study concerns non-linear ordinary differential equations of 
second order of the form 
\begin{equation}\label{eq}
y y''=F(z,y^2)~, ~~y: \R^+\to \R^+ ,
\end{equation}
where, as usual, the prime sign denotes derivative  with respect to the independent 
variable and $F$ is a suitably regular given 
function of its arguments. A large number of differential equation which model physical phenomena 
can be written in the form (\ref{eq}): we can mention the Ermakov-Pinney equation  
\cite{Common-Musette},\cite{Ermakov} and the Emden-Fowler 
equation (see e.g. \cite{Cha} and \cite{GH}).  These two equations and their generalisations
 are further discussed  in the subsequent sections. Finally, note that,  when 
the function $F$ in (\ref{eq}) depends on the independent $z$ variable only,   equation (\ref{eq}) 
reduces to the ordinary differential equation (1.1)  in \cite{JMAA2000}, which takes its physical 
origin in extended kinetic theory   (see \cite{JMAA2000} and references therein).   

It is well known, see \cite{Hille} for instance,  that the Schwarzian derivative plays a fundamental 
role in the theory of \emph{linear} second 
order differential equations. Given any smooth enough function $f(z)$ defined on an open set $I\subset\R$, 
its Schwarzian derivative $\{f,z\}$ is defined via 
\begin{equation}\label{Schw}
\{f,z\}:=\frac{f'''}{f'}-\frac{3}{2}\left(\frac{f''}{f'}\right)^2, ~~~ f'(z) \neq0~ ~~\forall z\in I.
\end{equation}
The link between the Schwarzian derivative and the theory of linear differential equations is given by the following result (see e.g. \cite{Hille}, Theorem 10.1.1): if $g_0$ and $g_1$ are two independent solutions of the equation
\begin{equation}\label{linear}
g''=2P' g'+Q g,
\end{equation}
where $P$ and $Q$ denote smooth functions, say $C^2$, of the independent variable, then the 
Schwarzian derivative of the ratio $g_0/g_1$, assuming $g_1\neq 0$,  depends only on the functions 
$P$ and $Q$. Specifically, it turns out \begin{equation}\label{pr}
\{\frac{g_0}{g_1},z\}=2(P''-P'^2-Q).
\end{equation}
 The latter  connects   equations  (\ref{linear}) and (\ref{eq}) via the Schwarzian 
derivative. In the following, B\"acklund transformations,  admitted by  (\ref{eq}) itself, are  obtained.
Then,   new explicit solutions of  (\ref{eq}) are  constructed via
B\"acklund transformations  admitted by  (\ref{eq}) itself.

The material is organised as follows. The opening Section 2 is concerned about the construction of a 
B\"acklund transformation  needed to establish the subsequent results. In particular, conditions which 
imply it is an auto-B\"acklund transformation are established. The functional equation which guaranties the B\"acklund 
transformation is an auto-B\"acklund transformation are studied in Section 3. 
In the following Section 4, the relation between the linear equation (\ref{linear}), the equation (\ref{pr}) 
and the auto-B\"acklund of the ordinary differential equations under investigation are considered.  
Notably, they exhibit  terms in which the Schwarzian derivative appears. 
 
 Section 5 is devoted to show how some well known 
 nonlinear ordinary differential equations are amenable to be treated via the method 
 devised in the previous Sections. Specifically, the physically relevant cases of the 
Ermakov-Pinney and of the Emden-Fowler equation are studied in two different subsections where  
corresponding   B\"acklund and auto  B\"acklund transformations are constructed.

In the closing Section 6, further to some conclusions, also perspective investigations are mentioned.
\section{B\"acklund transformations}
In this Section an invariance shown in \cite{JMAA2000} is revisited and suitably generalised to 
adapt to the wider class of nonlinear ordinary differential equations under investigation. To start with, 
let us introduce the following Lemma.

\begin{lemma}\label{lemma1} {Consider the ordinary differential equation \eqref{eq}, 
where $y: \R^+\to \R^+$,
then admitted solutions are related, via  B\"acklund transformation, to solutions of the 
ordinary differential equation
\begin{equation}\label{2}
v'' v -\frac{1}{2}(v')^2=2v F(z, v).
\end{equation}
}
\end{lemma}
 
 \noindent\textbf{Proof} Apply the transformation
\begin{equation}\label{v}
T: ~~~~ v= y^2~~, ~~ y, v: \R^+\to \R^+,
\end{equation}
and note that the kernel of $T$ is empty since $y$ and $v$ are assumed to be real positive valued. 
Then, direct substitution of \eqref{v} in  \eqref{eq} gives \eqref{2}.

 \hfill $\square$

\noindent
The link can be graphically depicted as follows
\begin{equation*}
\boxed{y''y=F(z,y^2)}{\buildrel T \over{-\!-}}\boxed{v'' v -\frac{1}{2}(v')^2=2v F(z, v)}
\end{equation*}

\begin{lemma}\label{lemma2} {Consider the reciprocal transformation 
\begin{equation}\label{reciprocal}
R:  ~~~~ \left\{\begin{array}{ccc}x = \Phi &  {D_t} = v {D_x},  \\ \\
  v =  D_t \Phi  & ~~~~~ D_x = [ \Phi_t]^{-1}{D_t}  \end{array}\right.
  \end{equation}
wherein:
\begin{equation}\label{R33}
\displaystyle{ {D_t} := {d  \over
dt},~~~~ {D_x} :=  {d  \over dx}, ~~~ \Phi_t = {d  \over dt} \Phi};
\end{equation}
it transforms  \eqref{2}, where $v: \R^+\to \R^+$, in 
\begin{equation}\label{eqphi}
\{\phi,t\}=2\phi_t F(\phi,\phi_t).
\end{equation}
}
\end{lemma}

\noindent\textbf{Proof} Direct application of the reciprocal transformation $R$ produces the result.

 \hfill $\square$
 
The two Lemmata can be summarised in the following   B\"acklund chart, according to the terminology in \cite{SIGMA2016}\footnote{The interested reader is referred to \cite{SIGMA2016} and Ref.s therein to track early occurences of the term {\it B\"acklund chart}.},  where by {\it B\"acklund chart} is meant the
net of links, represented by B\"acklund Transformations, among different equations.
\begin{equation*}
\boxed{y''y=F(z,y^2)}{\buildrel T \over{-\!-}}\boxed{v'' v -\frac{1}{2}(v')^2=2v F(z, v)}{\buildrel R \over{-\!-}}\boxed{\{\phi,t\}=2\phi_t F(\phi,\phi_t)}
\end{equation*}
Hence, a correspondence between equations  (\ref{eq}) and  (\ref{eqphi}), we term
 {\it   Schwarzian} equation, is established.  This result, combined with known properties 
of the Schwarzian derivative, allows to prove further invariances  enjoyed by  (\ref{eq}). 

\begin{propn}\label{prop2} {Let $g$ and $\psi$ denote two suitably smooth maps, then 
assume $\phi$  can be expressed via the composition of such two maps, namely $\phi=g(\psi)$,
then $\psi$ satisfies a Schwarzian equation of the same form of  (\ref{eqphi}), that is
\begin{equation}\label{eqphi2}
\{\psi,t\}=2\psi_t \tilde F(\psi,\psi_t)~,~ ~~\text{where}~ \tilde F~ \text{is a smooth function}.
\end{equation}}
\end{propn}

\noindent\textbf{Proof} Computation of the Schwarzian  derivative $\{\phi,t\}$, where  $\phi=g(\psi)$,
composition of the two differentiable maps $g$ and $\psi$, gives
\begin{equation}\label{Schwarzcomp}
\{\phi,t\}=\psi_t^2\{g,\psi\}+\{\psi,t\}.
\end{equation}
The latter, substituted in equation (\ref{eqphi}) can be re-written as
\begin{equation}
\{\psi,t\}=2\phi_tF(\phi,\phi_t)-\psi_t^2\{g,\psi\}.
\end{equation}
Note that this equation is of the same form of (\ref{eqphi}). Indeed, the derivative $\phi_t$ is 
$ g'\psi_t$, and hence 
\begin{equation}\label{psi}
\{\psi,t\}=2g'\psi_tF(g,g'\psi_t)-\psi_t^2\{g,\psi\}.
\end{equation}
Thus, if we define the new function $\tilde{F}$ via
\begin{equation}
\tilde{F}(\psi,\psi_t):= g'F(g,g'\psi_t)-\frac{\psi_t}{2}\{g,\psi\},
\end{equation}
and substitute it into equation (\ref{psi}), the thesis readily follows since equation  (\ref{eqphi2}), 
of the form of (\ref{eqphi}), is obtained.
 \hfill $\square$

Now,  the B\"acklund chart  can be extended  setting $B_1: \phi-g(\psi)=0$, so that the links are summarised 
as follows.
\begin{equation*}
{\small{\boxed{y''y=F(z,y^2)}{\buildrel T \over{-\!-}}\boxed{v'' v -\frac{1}{2}(v')^2=2v F(z, v)}{\buildrel R \over{-\!-}}\boxed{\{\phi,t\}=2\phi_t F(\phi,\phi_t) }{\buildrel {B_1} \over{-\!-}}\boxed{\{\psi,t\}=2\psi_t \tilde F(\psi,\psi_t) }}}
\end{equation*}

Accordingly, solutions admitted by the differential equation (\ref{eq}) can be constructed as stated in the following proposition.

\begin{propn}\label{prop3} {Let $f$  denote a smooth enough function such that $f'(z)\neq 0$, $\forall z\in\R^+$
introduce 
\begin{equation}\label{map}
Y^2(z):=\frac{y^2(f(z))}{f'(z)},
\end{equation}
where $y(z)$ is any solution of the equation (\ref{eq}),
then $Y(z)$ is a solution of the equation
\begin{equation}\label{te}
YY''=f'F(f,f'Y^2)-\frac{1}{2}\{f,z\}Y^2~.
\end{equation}
}
\end{propn}

\noindent\textbf{Proof} Note that \eqref{map} can be regarded as the  B\"acklund  transformation:
\begin{equation}\label{map2}
B_2: ~~{f'(z)} Y^2(z)-{y^2(f(z))}=0
\end{equation}
which, given the smooth function $f$, connects \eqref{eqphi2} to \eqref{eq}. Thus, esplicit computations 
prove the result.

 \hfill $\square$

\noindent

Let us discuss this result. The transformation (\ref{map}) represents a map between corresponding 
solutions of equations (\ref{eq}) and  (\ref{te}). Two different cases can occur. \\
That is,  when the two equations (\ref{te}) and  (\ref{eq}) exhibit different forms, the B\"acklund 
transformation $B_2$, given by  (\ref{map2}), connects solutions of equation (\ref{eq}) with solutions 
of equation (\ref{te}). \\
A special case arises when equations (\ref{eq}) and (\ref{te}) share the same form: in this case, $B_2$, 
given by (\ref{map}),  represents an auto-B\"acklund transformation between solutions of equation 
(\ref{eq}). The necessary and sufficient condition $B_2$ must satisfy to be an auto-B\"acklund 
transformation reads
\begin{equation}\label{fde}
F(z,v)=f'F(f,f'v)-\frac{1}{2}\{f,z\}v.
\end{equation}
The latter is  obtained via direct comparison between the r.h.s.s of the two  equations  (\ref{te}) and  
(\ref{eq}). This result is stated in the following corollary.
\begin{cor}\label{cor1}
Let $y(z)$ and $Y(z)$ be, in turn, solutions of the differential equations \eqref{eq} and  (\ref{te}), then 
$B_2$, in  \eqref{map}, represents an auto-B\"acklund transformation, whenever $f$ denotes a solution of 
the functional differential equation \eqref{fde}.
\end{cor}
Equation \eqref{fde} is a functional differential equation: a characterisation of the solutions it admits  is provided
in the next section. Then,  explicit examples
are considered to show how it is possible to construct new solutions of equation (\ref{eq}) on
application of the transformation (\ref{map2}).

\section{The functional differential equation} 

This Section is concerned about the study of the functional equation (\ref{fde}) wherein the unknown 
is the function $f$, while $F$ is given. Thus, it is a 3rd order nonlinear ordinary differential equation 
subject to the condition $f^{\prime}\neq 0$.
Here, for the sake of simplicity, we do not look for  $f$, which is a solution of  (\ref{fde}), but   we prefer to 
consider (\ref{fde})  as a linear non homogeneous ordinary differential equation in the unknown  $F$, where, 
in addition, $F$ is assumed to depended on $v$ via a {\it suitable} power expansion. This approach allows 
us to say which forms the given  function $F$ may assume in order  to be amenable to the presented method. 
Specifically, given a suitable $F$, a method to construct solutions of equation \eqref{eq}, via  B\"acklund 
transformations, is provided.

Accordingly, let us assume that the function $F(z,v)$  admits a formal power expansion in $v$, that is 
\begin{equation}\label{Fexp}
F(z,v)=\sum_{n}F_{n}(z)v^n,
\end{equation}
where the sum is intended on a suitable set of integers. For the sake of convenience, when $n\neq 1$, the 
coefficients $F_n(z)$ are looked for under the form $F_{n}=(G'_n)^{n+1}$, where $G'_n$ denotes the 
derivative of a smooth (at least $C^1$) function of $z$. In the case $n=1$, we set 
\begin{equation}\label{F1}
F_1(z):=(G'_1)^{2}-\frac{1}{2}Q(z)
\end{equation}
where $Q(z)$ is a suitable function. Hence, substitution of the latter and of  \eqref{Fexp} in (\ref{fde}),  gives 
\begin{equation}\label{fde1}
2\sum_{n}\left((f')(G'_n)(f)-(G'_n)(z)\right)^{n+1}v^n-(f')^{2}Q(f)v+Q(z)v-\{f,z\}v=0.
\end{equation}
which depends explicitly on $G_n$, so that, for example, $(G'_n)(f)$ indicates that  $G'_n$ is composed with $f(z)$.
 \\
The power series expansion in (\ref{fde1}) represents a polynomial in $v$ whose first term multiplies $v^{-1}$,  hence equality (\ref{fde1}) is satisfied when all the coefficients of  $v^{k}$ are set 
equal to zero. This means that the action of the map $f(z)$ on the functions $G_n(z)$ is a translation, 
that is, the functions $G_n(z)$ satisfy the linear functional equations
\begin{equation}\label{G_nK}
G_n(f(z))=G_n(z)+K_n~~~\Longleftrightarrow~f'G'_n(f)=G'_n(z)~~\forall n,
\end{equation}
where $K_n$ are arbitrary constants.  Substitution of \eqref{G_nK} in (\ref{fde1}) gives
\begin{equation}\label{fde2}
Q(z)=(f')^{2}Q(f)+\{f,z\}.
\end{equation}
Notably, the right hand side of the latter reminds the right hand side of equation (\ref{Schwarzcomp})  when, in turn,  the variable $z$ is identified with the variable $t$ and the function $f$ with the function $\psi$. This correspondence suggests to identify $Q$ with a suitable Schwarzian derivative, that is, to set
\begin{equation}\label{position}
Q(z):=\{w,z\}.
\end{equation}
Substitution of this position in \eqref{G_nK}, and the subsequent comparison of \eqref{fde2} with  
(\ref{Schwarzcomp}) show that solutions of \eqref{fde2}  are obtained whenever  $w(z)$ and $f(z)$ are 
related via the functional equation
\begin{equation}\label{position2}  
\{w(z),z\}=(f')^{2}\{w(f),f\}+\{f,z\}, %~~\text{wherein}~~w(z)\equiv w(f(z))~\text{e.i.}~\{w(z),z\}=\{w(f(z)),z\},
\end{equation}
wherein the Schwarzian derivative of the composition of two functions appears since 
$(f')^{2}\{w(f),f\}+\{f,z\}\equiv \{w(f),z\}$. Accordingly, the function $w(f(z))$ and $w(z)$ follow to be related via a fractional linear transformation, i.e.
\begin{equation}\label{flt}
w(f(z))=\frac{aw(z)+b}{cw(z)+d}, \qquad ad-bc\neq 0.
\end{equation}
Hence, combination of the shown results, allows to prove the following proposition.
\begin{propn}\label{prop1}
If the function $F(z,v)$ can be represented as a power series expansion in $v$, then the solution 
of the functional differential equation
\begin{equation*}
F(z,v)=f'F(f,f'v)-\frac{1}{2}\{f,z\}v.
\end{equation*}
is given by
\begin{equation}\label{F}
F(z,v)=\sum_n a_n \left(G'_n (z)\right)^{n+1}v^n-\frac{1}{2}\{w(z),z\}v
\end{equation}
where the functions $G_n(z)$ and $w(z)$ satisfy the functional equations ($a$, $b$, $c$, $d$ and $K_n$ are arbitrary constants)
\begin{equation}\label{functionaleqs}
G_n(f(z))=G_n(z)+K_n, \quad w(f(z))=\frac{aw(z)+b}{cw(z)+d}, \qquad ad-bc\neq 0.
\end{equation}
\end{propn}
\smallskip
The previous proposition allows to find differential equations possessing the auto-B\"acklund transformations 
(\ref{map2}).  A corollary of this proposition reads as  follows.
\begin{cor}\label{cor1}
Let $y_0(z)$ be a solution of the differential equation
\begin{equation}\label{diffeq}
y''=\sum_n a_n \left(G'_n (z)\right)^{n+1}y^{2n-1}-\frac{1}{2}\{w(z),z\}y,
\end{equation}
where the functions $G_n(z)$ and $w(z)$ are specified in Proposition \ref{prop1}, then 
$$
y_1(z)^2=\frac{y_0(f(z))^2}{f'(z)}
$$
represents an auto-B\"acklund transformation admitted by equation (\ref{diffeq}). 
\end{cor}

\noindent These results can be enriched and complemented thanks to the following two propositions.
\begin{propn}\label{propnG}
The general solution of the functional equation 
\begin{equation}\label{fe1}
G(f(z))=G(z)+K
\end{equation}
is given by
\begin{equation}\label{sfe1}
G(z)=G_0(z)+\Phi(G_0(z))
\end{equation}
where $G_0(z)$ is a particular solution of equation (\ref{fe1}) and $\Phi(z)$ is 
an arbitrary periodic function of period $K$.
\end{propn}

\begin{propn}\label{propnw}
Let $w(z)$ be defined as 
\begin{equation}\label{sfe2}
w(z):=\frac{DG(z)-B}{A-CG(z)}~, ~~ 	A, B, C, D \in \R
\end{equation}
where $G(z)$ is any function satisfying equation (\ref{fe1}), then $w(z)$ is a solution of 
\begin{equation}\label{fe2}
w(f(z))=\frac{(\Delta+DCK)w(z)+D^2K}{(\Delta-DCK)-w(z)C^2K}.
\end{equation}
where $\Delta = AD-BC\neq0$.
\end{propn}
These propositions can be easily proved via direct computations.

\section{Auto-B\"acklund transformations }
 
In this Section auto-B\"acklund transformations are related to linear  
second order ordinary differential equations. Specifically, the attention is focussed on  how   solutions 
admitted by second order \emph{linear} ordinary differential equations lead to the construction of  the transformation (\ref{map2}).

First of all, the general ideas are presented. Consider the link between two 
independent solutions admitted by  (\ref{linear}) and let $w(z)$ denote their ratio, namely
\begin{equation}\label{w}
w(z):= \frac{g_0(z)}{g_1(z)}, ~~ g_1(z)\neq 0.
\end{equation}
Then, according to  (\ref{pr}), the Schwarzian derivative of $w$  is given by
$$
-\frac{1}{2}\{w,z\}=Q+(P')^2-P'',
$$
where $P$ and $Q$ are the smooth coefficients of the ordinary differential equation  (\ref{eq}). 
Assume, now, a function $f(z)$ exists  such that $w(f(z))$ can be expressed 
as a fractional linear transformation of $w(z)$, i.e. it is such that it satisfies the 
hypotheses of proposition \ref{prop1}. Consider, then, the fractional linear 
transformation: 
\begin{equation}\label{wf}
w(f(z))=\frac{(\Delta+DCK)w(z)+D^2K}{(\Delta-DCK)-w(z)C^2K}~, ~~ 
	A, B, C, D, K \in \R
\end{equation}
where $\Delta=AD-BC\neq0$. Let the function $G(z)$ be defined via
$$
G(z):=\frac{Aw(z)+B}{Cw(z)+D},
$$
then, the functional relation  $G(f(z))=G(z)+K$  holds, $\forall z$. Note that the 
function $w(z)$ is given by (\ref{w}); hence,  the derivative of the function $G(z)$ 
can be written, on use of the Wronskian  of solutions of equation (\ref{linear}), as 
\begin{equation}\label{Gprime}
G'(z)=\Delta C_1 \left(\frac{e^{P(z)}}{Cg_0(z)+Dg_1(z)}\right)^2,
\end{equation}
where $C_1$  is a suitably chosen  constant. When the 
arbitrary constants, denoted as $a_n$, are inserted as coefficients, 
the following, consequence of Corollary \ref{cor1} is proved.
\begin{cor}\label{cor2}
Let $y_0(z)$ be a solution of the differential equation
\begin{equation}\label{diffeq1}
y''=\sum_n a_n \left(\frac{e^{P(z)}}{Cg_0(z)+Dg_1(z)}\right)^{2n+2}y^{2n-1}+\left(Q+(P')^2-P''\right)y,
\end{equation}
where the functions $P(z)$ and $Q(z)$ are arbitrary and $g_0$ and $g_1$ are two independent 
solutions of the linear differential equation (\ref{linear}). Then, if it is possible to find a function 
$f(z)$ such that equation (\ref{wf}) holds, the map
\begin{equation}\label{BTs}
y_1(z)^2=\frac{y_0(f(z))^2}{f'(z)}
\end{equation}
defines an auto-B\"acklund transformation admitted by equation (\ref{diffeq1}). 
\end{cor}

The relevance of this result is stressed in the next section where  the results 
(\ref{diffeq1})-(\ref{BTs}) are  applied to different examples.

\section{Applications}
The class of differential equations  (\ref{diffeq}) can be shown to embrace, as particular cases, a
large number of nonlinear differential equations. The aim of this Section is to show implications
on the well known Ermakov-Pinney and Emden-Fowler nonlinear equations whose applicative
 relevance is undoubtable. In addition, some generalisations of these equations are also 
 considered.

 First of all, a simple case is treated. Note that if,  in  (\ref{diffeq}), here re-written for convenience
$$
y''=\sum_n a_n \left(G'_n (z)\right)^{n+1}y^{2n-1}-\frac{1}{2}\{w(z),z\}y, \eqno(29)
$$ 
 we set $G_n(z):=G(z), ~\forall n$, it is compatible with the further positions $K_n = K$   and $\Phi =0$.  
Indeed, in \eqref{functionaleqs}  the arbitrariness in the choice of the functions 
$G_n(z)$  is reflected in the arbitrariness of the parameters $K_n$ and of the periodic function $\Phi$, 
respectively, in equation (\ref{G_nK}) and  (\ref{sfe1}).
Furthermore,  the (tacitly) assumed convergence, in a given open set $\Omega\subset\R$, of the series in  
(\ref{diffeq}), implies that the sequences $\{a_n\}\in \R$ is such that, in $\Omega\subset\R$,
the following convergent series can be introduced 
\begin{equation}\label{H}
H(z):=\sum_{n} a_n z^n.
\end{equation}
The Corollary \ref{cor1} takes, in this case, the following simplified form.
\begin{cor}\label{cor4}
Let $y_0(z)$ be a solution of the differential equation
\begin{equation}\label{diffeq3}
yy''=G'H(G'y^2)-\frac{1}{2}\{w,z\}y^2,
\end{equation}
where the functions $G(z)$ and $w(z)$, as specified in  Proposition \ref{prop1},
 satisfy the functional equations \eqref{functionaleqs}, that is 
 \begin{equation*}
G(f(z))=G(z)+K, \quad w(f(z))=\frac{aw(z)+b}{cw(z)+d}, ~~~a,b,c,d, K \in\R.  \hskip3cm
(\ref{functionaleqs})
\end{equation*}
 Then, the map
$$
y_1(z)^2=\frac{y_0(f(z))^2}{f'(z)}
$$
defines an auto-B\"acklund transformation of the equation (\ref{diffeq3}). 
\end{cor}
In the next subsections, differential equations with relevant physical meaning are obtained 
corresponding to {\it ad hoc} choices, in  (\ref{diffeq3}),  of the  function $H(z)$.

\subsection{The Ermakov-Pinney equation}
A first remarkable case is represented by the Ermakov-Pinney equation studied in this subsection.
Specifically, if, in the equation (\ref{diffeq1}), the function  $P(z)$ as well as all the coefficients 
$a_n$  are  all set equal to zero except $a_{-1}$,  denoted via  $a_1=-\alpha$,
then, the Ermakov-Pinney equation \footnote{On introduction of the 
Kronecker symbol ${\displaystyle{\delta_{k, l}:= \left\{ { \begin{array}{c}
    0 ~~ k\neq l \cr
    1~~ k = l \\ 
  \end{array} } \right.}}$, the condition on the coefficients $a_n$ can be
  written as  $a_n= -\alpha \delta_{-1, n}$. }: 
\begin{equation}\label{Pinney}
y''=Q(z)y-\frac{\alpha}{y^3} 
\end{equation}  
is obtained. This is equivalent to choose $H(z)=-\frac{\alpha}{z}$ in corollary \ref{cor4}. 

The Ermakov-Pinney equation was introduced in 1880 by Ermakov \cite{Ermakov}, who investigated
solvability conditions of  second order ordinary differential equations. The Ermakov-Pinney equation  
is closely connected to the harmonic oscillator 
with a time-dependent frequency. It is used to describe various problems,  in quantum 
mechanics, such as  the motion of charged particles in the Paul trap \cite{MGW}, in atomic 
transport  theory  \cite{T} and  Bose-Einstein condensate theory \cite{Nicolin}. Furthermore,
 it plays a fundamental role in the description of the unitary evolution of quantum non-
 autonomous systems \cite{VV}. It appears also  in cosmology \cite{HL}. A set of B\"acklund 
 transformation and  an exact discretization admitted by  equation (\ref{Pinney}) is due to
  A. Hone \cite{AH}, who, on    application of two different  transformations, provides its 
  exact discretization.   \\
Now, on applications of  results in the previous sections,  a set of  B\"acklund transformations 
admitted by the Ermakov-Pinney equation are obtained. In particular, since equation  
(\ref{Pinney}) corresponds to  $P=0$ in equation (\ref{diffeq1}), the  admitted B\"acklund 
transformations are specified by the following proposition.
\begin{propn}\label{PinneyBT}
Suppose that $y_0$ is a solution of the Ermakov-Pinney equation. Define the function $\displaystyle{w(z)=\frac{g_0}{g_1}}$, where $g_0$ and $g_1$ are two independent 
solutions of the linear differential equation $y''=Qy$. Then, if it is possible to find a function 
$f(z)$ such that equation (\ref{flt}) holds, the map
\begin{equation*}
y_1(z)^2=\frac{y_0(f(z))^2}{f'(z)}
\end{equation*}
is an auto-B\"acklund transformation of the equation (\ref{Pinney}). 
\end{propn}

\noindent\textbf{Proof} The proposition follows on specialisation, in Corollary \ref{cor2}, 
of all the parameters via $a_k=-\alpha\,\delta_{k,-1}$ and $ P\equiv 0$.

\hfill$\square$

Notable examples are represented by the case when  the function $Q$   exhibits a dependence 
on  $z^2$: in particular, it is a linear combination of a term proportional to $z^{-2}$ with a term 
proportional to $z^{4k}$ for some integer $k\in\N$. Both these functions, taken separately, have 
application in scalar field cosmologies (see \cite{HL}, eqs. (10)-(11)-(31)-(38)).  
Specifically, consider $Q$ to be
\begin{equation}\label{ex1}
Q(z)=p^2(2k+1)^2\frac{z^{4k}}{4}+\frac{k(k+1)}{z^2} ~~, ~ p\in\R^+.
\end{equation}
In this case, the function $w$ is obtained 
\begin{equation}\label{wex1}
w(z)=e^{pz^{2k+1}}
\end{equation}
and the function $f(z)$ can be defined by
\begin{equation}\label{fex1}
f^{2k+1}=\frac{1}{p}\ln\left(\frac{aw(z)+b}{cw(z)+d}\right)~~, ~ ad-bc\neq0.
\end{equation}
Hence, given  a solution $y_0$ of equation (\ref{Pinney}), further solutions are represented by
\begin{equation}\label{BTex1}
y_1(z)^2=\frac{y_0(f)^2}{f'}.
\end{equation}
Indeed, chosen $Q$ in  (\ref{ex1}), a particular solution of equation (\ref{Pinney}) 
 is given by
\begin{equation}\label{psex1}
y_0(z)=\frac{\beta}{z^k}, \qquad \beta^4:= \frac{4\alpha}{p^2(2k+1)^2}.
\end{equation}
Thus, on application of  the transformation (\ref{BTex1}), after few manipulations, 
we find a new set of solutions, say $y_1(z)$, as follows 
\begin{equation}\label{solex1}
y_1(z)^2=\frac{\beta^2}{z^{2k}}\frac{(aw(z)+b)(cw(z)+d)}{w(z)(ad-bc)}~~, ~ ad-bc\neq0~,
\end{equation}
where $w(z)$ is defined by equation (\ref{wex1}). Note that the previous equation defines 
the general solution of equation (\ref{Pinney}) which corresponds to $Q$  in (\ref{ex1}).

\subsection{The Emden-Fowler equation}
This subsection is devoted to a second example of an ordinary differential equation 
which admits B\"acklund transformations whose construction is possible according to 
the general results comprised in Sections 2-4. The generalised Emden-Fowler equation 
of the first kind in the unknown $q(x)$ reads (see e.g \cite{GH})
\begin{equation}\label{EF1}
xq_{xx}+\alpha q_{x}+\beta x^{\nu}q^n=0.
\end{equation}
This equation appears in a wide variety of mathematical physics problems: specifically, 
when $\nu=1$ and $\alpha=3$ it describes the hydrostatic and thermodynamic equilibrium 
of stars. In addition, it appears in  investigations on the Einstein's field equations \cite{Hertl} 
or in the mean-field description of critical adsorption \cite{GR} 
(see \cite{GH} and references therein for further examples). When the parameter  
$\alpha$ is integer, equation (\ref{EF1})  is used to model  spherically symmetric steady 
state solutions of evolution problems involving the Laplace operator in a 
$\alpha$-dimensional space. More precisely, when $r$ denotes the {\it radial} coordinate,
looking for  spherically symmetric solutions,  the reduction of the equation
$$
\nabla^2 q+\beta r^{\nu-1}q^n=0
$$
directly gives equation (\ref{EF1}). A particularly interesting generalisation is represented by  
the modified Emden-Fowler equation, which is obtained on substitution of  the factor $x^{\nu}$, 
in (\ref{EF1}),  with a function of $x$:
\begin{equation}\label{EFM}
xq_{xx}+\alpha q_x+\beta r(x)q^n=0.
\end{equation}
In this section  two examples are provided: one for equation (\ref{EF1}) and the other for equation 
(\ref{EFM}).  Equation (\ref{EF1}) is the particular case of equation (\ref{EFM}), which corresponds 
to the choice $r(x)=x^{\nu}$. Likewise, our first example is a particular case of the second one. \\
\textbf{Example 1}. In this example, let  $\nu=1-2\alpha$ in  (\ref{EF1}) and, for later convenience, 
 set $n=2m-1$. The differential equation (\ref{EF1}) then reads
\begin{equation}\label{qeq}
x q_{xx}+\alpha q_x +\beta x^{1-2\alpha} q^{2m-1}=0.
\end{equation}
The following change of variables 
$$
q(x)=\frac{y(z)}{z}, \qquad z=x^{\alpha-1}
$$
in  (\ref{qeq}) reduces to
\begin{equation}\label{EF1R}
y''+\frac{\beta}{(\alpha-1)^2} \frac{y^{2m-1}}{z^{2m+2}}=0.
\end{equation}
The latter can be compared with  equation (\ref{diffeq1}): that is, according to  Corollary \ref{cor2}, 
when we set  $Q=P''-(P')^2$ and $a_m=- \frac{\beta}{(\alpha-1)^2}\delta_{m,n}$,   equation 
(\ref{EF1R}) is obtained. Indeed, 
the choice  $Q=P''-(P')^2$ entails that the functions $g_0$ and $g_1$ are linear combinations of 
$e^P$ and $ze^P$ (independent solutions of equation (\ref{linear})). Let, now chose $$
g_0=e^P(D-bz),\qquad g_1=e^P(az-C), \qquad aD-bC=1
$$
then, equation (\ref{diffeq1}) reads exactly as (\ref{EF1R}). The next step  is to show, 
solving the functional equation (\ref{fe2}), that the function $f(z)$   is given by
$$
f(z)=\frac{\Delta z}{Kz+\Delta},
$$
which allows to find the B\"acklund transformations
\begin{equation}\label{BTexx1}
y_1^2(z)=y_0^2\left(\frac{\Delta z}{Kz+\Delta}\right)\frac{(Kz+\Delta)^2}{\Delta^2}.
\end{equation}
A particular solution of equation (\ref{EF1}) is given in \cite{GH}: it corresponds to the 
case when 
\begin{equation}\label{firstsol}
{\displaystyle{y_0(z)=p \, z^{\frac{m}{m-1}} }}
\end{equation}
and represents a solution of equation (\ref{EF1R}), where the constant $p$ is required to
 satisfy
$$
\beta (m-1)^2p^{2m-2}+m(\alpha-1)^2=0.
$$
Note  that the latter, in general, given  real  $\beta$ and $m$, admits complex solutions 
$p\in\C$; hence, to obtain $p\in\R$,  $\beta$ and $m$ are required to be, in turn, positive 
and negative, or viceversa.  The new solution (\ref{BTexx1}) reads
\begin{equation}\label{firstbac}
y_1^2(z)= p^2\left(\frac{\Delta z}{Kz+\Delta}\right)^{\frac{2m}{m-1}}\frac{(Kz+\Delta)^2}{\Delta^2}
\end{equation}
and represents two different one-parameter solutions of equation (\ref{EF1R}), the parameter we can chose 
is $\frac{K}{\Delta}$.  Note also that further iterations of the same transformation, where  each iteration is
identified by the pair of parameters $\Delta_n$ and $K_n$, do not add further  constants 
to the solution. Indeed, the $n^{th}$ iteration can be written as
$$
y_n^2(z)=p^2  \left(\frac{R_n z}{S_nz+R_n}\right)^{\frac{2m}{m-1}}\frac{(S_n z+R_n)^2}{R_n^2}
$$
where the coefficients $R_n$ and $S_n$ solve the recurrence relations
$$
 R_{n+1}=\Delta_{n+1}R_n,\quad S_{n+1}=\Delta_{n+1}S_n+K_{n+1}R_n
$$
with the initial values $R_0=1$, $S_0=0$ (and obviously $\Delta_1=\Delta$ and $K_1=K$).\\
\newline
\textbf{Example 2}. In this example we look at the modified Emden-Fowler equation (\ref{EFM}). Again, 
for later convenience, we set $n=2m-1$. We specify the function $r(x)$ as
\begin{equation}\label{setr}
r(x)=\frac{1}{x^{2\alpha}}\left(\frac{x^{\alpha-1}}{\eta+\gamma x^{\alpha-1}}\right)^{2m+2},
\end{equation}
which corresponds to the following differential equation in the unkown $q(x)$
\begin{equation}\label{EFMq}
xq_{xx}+\alpha q_x+\beta x^{1-2\alpha}\left(\frac{x^{\alpha -1}}{\eta+\gamma x^{\alpha-1}}\right)^{2m+2} q^{2m-1}=0.
\end{equation}
When the special values $(\eta,\gamma)=(0,1)$ are chosen,   (\ref{EFMq}) reduces to 
 (\ref{qeq}) of the previous example. The subsequent change of variables in (\ref{EFMq}) 
$$
q(x)=\frac{y(z)}{z}, \qquad z=x^{\alpha-1},
$$
gives 
\begin{equation}\label{yr}
y''+\frac{\beta}{(\alpha-1)^2}\frac{y^{2m-1}}{(\eta+\gamma z)^{2m+2}}=0.
\end{equation}
This equation can be compared with  equation (\ref{diffeq1}): according to  Corollary \ref{cor2}, we must set 
$Q=P''-(P')^2$ and $a_m=- \frac{\beta}{(\alpha-1)^2}\delta_{m,n}$.
 Then, the functions $g_0$ and $g_1$ are linear combinations of $e^P$ and $ze^P$ 
 (that is, they are two independent solutions of equation (\ref{linear})). Let 
\begin{equation*}
\left\{\begin{split}
g_0&=e^P(a_0 z+b_0),\\
g_1&=e^P(a_1z+b_1).
\end{split}\right.
\quad \textrm{with}\quad 
\left\{\begin{split}
Ca_0+Da_1=\gamma,\\
 Cb_0+Db_1=\eta,
\end{split}\right.
\end{equation*}
 These choices  allow to check that equation (\ref{diffeq1}) reduces exactly to (\ref{yr}). 
 Then it can be  shown that the function $f(z)$, solution of the functional equation (\ref{fe2}), is given by
$$
f(z)=\frac{(K\eta\gamma +\Delta\delta)z+K\eta^2}{(\Delta\delta-K\eta\gamma)-K\gamma^2z}, \qquad \delta:= a_0b_1-a_1b_0.
$$
According to Corollary \ref{cor2}, the B\"acklund transformations admitted by equation (\ref{yr}) are 
\begin{equation}\label{secondbac}
y_1(z)^2=y_0\left(\frac{(K\eta\gamma +\Delta\delta)z+K\eta^2}{(\Delta\delta-K\eta\gamma)-K\gamma^2z}\right)^2\left(\frac{(\Delta\delta-K\eta\gamma)-K\gamma^2z}{\Delta\delta}\right)^2.
\end{equation}
A particular solution of (\ref{EFMq}) is given by
\begin{equation}\label{secondsol}
y_0(z)=p(\eta+\gamma z)^{\frac{m}{m-1}}
\end{equation}
where the constant $p$ represents a solution of
$$
\beta (m-1)^2p^{2m-2}+m\gamma^2 (\alpha-1)^2=0.
$$
Note that the solution (\ref{secondsol}) reduces to (\ref{firstsol}) when 
$(\eta,\gamma)=(0,1)$ and, then, the transformation (\ref{secondbac}) coincides with 
the transformation (\ref{firstbac}).\\
The new solution is given by
$$
y_1(z)^2=p^2\left(\frac{\Delta\delta(\eta+\gamma z)}{(\Delta\delta-K\eta\gamma)-K\gamma^2z}\right)^{\frac{2m}{m-1}}\left(\frac{(\Delta\delta-K\eta\gamma)-K\gamma^2z}{\Delta\delta}\right)^2
$$

\section{Conclusions}
%{\bf{\textcolor{green}{Ho tolto le parti  $y: \R^+\to \R^+ $ ed $ S: \R^+ \to \R^+$ che qui mi sembrano inessenziali. Ho messo qui anche la parte che dicevi, cio\'e ``In particular, in the very special
%case when F depends only on z...''  {{\textcolor{red}{\bf O.K.}}} }}}\\
A class of nonlinear differential equations of second order of the form  $y y''=F(z,y^2)$, which represents a 
genuine generalisation of $y~ y^{\prime\prime} = S(x),$ in \cite{JMAA2000}, is considered in this article. The 
studied equations are proved to admit B\"acklund transformations  (\ref{map}) which, in the general case, 
represent maps from solutions of a nonlinear differential equations of the form (\ref{eq}) to solutions of another 
equation of  the same class. Notably,  these B\"acklund transformations comprise also, as special cases, 
auto-B\"acklund transformations. These special cases are investigated; thus, whenever $F(z,v)$ and $f(z)$ 
satisfy the   differential functional equation (\ref{fde}), it is shown  that  the map (\ref{map2}) represents an 
auto B\"acklund transformations admitted by equation (\ref{eq}). These auto B\"acklund transformations are 
algebraic and, indeed, in the complex domain, they establish multi-valued correspondences among solutions 
of the same differential equations; however, when restrictions to  positive  real valued functions are 
considered, one-to-one correspondences are obtained.
In particular, in the very special case when $F$ depends only on $z$, i.e. $F(z,v)=S(z)$, and,
 furthermore, the function $f$ is a linear fractional transformation, i.e. $f(z)=(az+b)/(cz+d)$, 
 the transformation  in \cite{JMAA2000} is obtained as a particular case of  (\ref{map2}).
In addition, a variety of applications of equation (\ref{eq}) are provided: they represents only some of the examples in which the constructed B\"acklund transformations can be applied to obtain new results. Further systems of physical interest, which, for instance arise as reductions of partial differential equations, belong  to the   class of equations we investigate: in the present article we restrict our attention to some physically relevant examples. Further results, in particular as far as  the structure  of equation (\ref{eq}) as well as its solutions in the complex domain are in progress. Also, currently under investigation are relations between the obtained  B\"acklund transformations,  integrability properties and, possibly, the Hamiltonian structure of the linked equations, aiming to specialise  the results in \cite{PhysA166}.

\begin{center} {\bf Acknowledgments} \end{center}
The financial  support of GNFM-INdAM, INFN and  \textsc{Sapienza} Universit\`a di Roma  is gratefully acknowledged.


\bigskip


\begin{thebibliography}{40}
\bibitem{Bob} A.I. Bobenko, B. Lorbeer, Yu B. Suris {\it Integrable discretizations of the Euler
top}, J. Math. Phys., {\bf 39}, (1998), 6668-6683.

\bibitem{JMAA2000} S. Carillo,  {\it A novel B\"acklund invariance of a nonlinear differential equation}, Journal of Mathematical Analysis and Applications,  {\bf 252}: 2, (2000), 828-839.
\bibitem{SIGMA2016}{  S. Carillo, M. Lo Schiavo, C. Schiebold},  {\it B\"acklund Transformations and Non Abelian Nonlinear Evolution Equations: a novel B\"acklund Chart},
Symmetry, Integrability and Geometry: Methods and Applications (SIGMA),  {\bf 12}, (2016), 087, 17 pages.
 % submitted 2015, arXiv:1512.02386, 15 pages
%\url{http://dx.doi.org/10.3842/SIGMA.2016.087}
%    doi \doi: 10.3842/SIGMA.2016.087
%\href{http://dx.doi.org/10.3842/SIGMA.2016.087}{http://dx.doi.org/10.3842/SIGMA.2016.087}
\bibitem{Cha} S. Chandrasekhar:  {\it An introduction to the study of stellar structures}, Dover, New York, 1958.
\bibitem{Common-Musette} A.K. Common,  M. Musette,
 {\it Two discretisations of the Ermakov-Pinney equation}
Physics Letters, Section A: General, Atomic and Solid State Physics,  {\bf  235}:6, (1997), 574-580. 
\bibitem{Ermakov} V. Ermakov:  {\it Second order differential equations. Conditions of
complete integrability}, Universita Izvestia Kiev Series III {\bf 9}, (1880), 1-25. English translation: A.O. Harin, 
Redactor: P.G.L. Leach, Applied Analysis and Discrete Mathematics,  {\bf 2} (2008), 123-145.
\bibitem{PhysA166}  B. Fuchssteiner, S. Carillo:   {\it
The Action-Angle Transformation for Soliton
Equations},  Physica A, {\bf 166}, (1990), 651-676.
%$ doi:10.1016/0378-4371(90)90078-7$
\bibitem{GR} S. Gbutzmann, U. Ritschel:  {\it Analytic solution of Emden-Fowler equation and critical adsorption
in spherical geometry}, Z. Phys. B, {\bf 96} (1995), 391-393.
\bibitem{GH} H. Goenner, P. Havas:  {\it Exact solutions of the generalized Lane–Emden equation}, Journal of Mathematical Physics, {\bf 41}, (2000), 7029.% doi: 10.1063/1.1308076
\bibitem{HL} R.L. Hawkins, J.E. Lidsey:  {\it Ermakov-Pinney equation in scalar field cosmologies}, Physical Review D, {\bf 66} (2002), 023523.
\bibitem{Hertl} E. Hertl:  {\it Spherically symmetric nonstatic perfect fluid solutions with shear}, General Relativity and Gravitation, {\bf 28}: 8 (1996), 919–934.
\bibitem{Hille} E. Hille,  {\it Ordinary differential equations in the complex domain}, Wiley, 1976, Toronto.
\bibitem{AH} A.N.W. Hone:  {\it Exact discretization of the Ermakov–Pinney equation}, Physics Letters A, {\bf 263} (1999), 347–354.
\bibitem{HKR} A.N.W. Hone, V.B.  Kuznetsov, O. Ragnisco,  {\it B\"acklund transformations for
many-body systems related to KdV}, J. Phys. A: Math. Gen., {\bf 32} (1999), 299-306.
\bibitem{HRZ} A. N. W. Hone, O. Ragnisco, F. Zullo,  {\it Algebraic entropy for algebraic maps}, J. Phys. A: Math. Theor. {\bf 49}: 2 (2016), 02LT01.
\bibitem{KPR} V.B. Kuznetsov, M. Petrera, O. Ragnisco,  {\it Separation of variables and  B\"acklund transformations for the symmetric Lagrange top}, J. Phys. A: Math. Gen.,
{\bf 37} (2004), 8495-8512.
\bibitem{KK} K. Kefalas:  {\it On smooth solutions of non linear dynamical systems $f_{n+1}=u(f_n)$, PART I}, Physics International {\bf 5}: 1 (2014), 112-127.% 2014. doi:10.3844/pisp.2014.112.127.
\bibitem{MGW} F. Major, V.N. Gheorghe, G. Werth,  {\it Charged Particle Traps - Physics and
Techniques of Charged Particle Field Confinement}, Springer, Berlin, 2005.
\bibitem{Moser} J. Moser, A.P. Veselov: {\it  Discrete versions of some classical integrable systems and
factorization of matrix polynomials}, Commun. Math. Phys., {\bf 139} (1991), 217-243.
\bibitem{Nicolin} A.I. Nicolin:  {\it Resonant wave formation in Bose-Einstein condensates}. Phys. Rev. E, {\bf 84} (2011), 056202.
\bibitem{Nijhoff} F.W. Nijhoff, O. Ragnisco, V.B. Kuznetsov:  {\it Integrable time-discretisation
of the Ruijsenaars-Schneider model}, Commun. Math. Phys., {\bf 176} (1996), 681-700.
\bibitem{Rag1} O. Ragnisco, F. Zullo,  {\it B\"acklund  transformations as exact integrable time discretizations for the trigonometric Gaudin model}, J. Phys. A, {\bf 43} (2010), 434029.
\bibitem{Rag2} O. Ragnisco, F. Zullo:  {\it B\"acklund transformation for the Kirchhoff top} Symmetry, Integrability and Geometry: Methods and Applications (SIGMA),
{\bf 7} (2011), 001, 13 pages.
\bibitem{SM} C. Small:  {\it Functional Equations and How to Solve Them}, Springer, 2007, New York. 
\bibitem{T} E. Torrontegui, S. Ibáñez, X. Chen, A. Ruschhaupt, D. Guéry-Odelin, and J. G. Muga: {\it Fast atomic transport without vibrational heating} Phys. Rev. A, {\bf 83} (2011), 013415.
\bibitem{Tsi1} A.V. Tsiganov:  {\it On Auto and Hetero B\"acklund Transformations for the Hénon–Heiles Systems}, Phys. Lett. A, {\bf 379} (2015), 2903–2907.
\bibitem{Tsi2} A.V. Tsiganov: {\it Integrable Euler Top and Nonholonomic Chaplygin Ball}, J. Geom. Mech., {\bf 3}: 3 (2011), 337–362.
\bibitem{VV} D.G. Vergel, E.J.S. Villasenor: {\it The time-dependent quantum harmonic oscillator revisited: Applications to quantum field theory}, Annals of Physics, {\bf 324} (2009), 1360–1385.
\bibitem{WW} E.T. Whittaker, G.N. Watson: {\it A course of modern analysis}, Cambridge University Press, 1927.
\bibitem{FZ1}  F. Zullo: {\it B\"acklund  transformations and Hamiltonian flows},  J. Phys. A,  {\bf 46}: 14, (2013), 145203.
\bibitem{FZ2}  F. Zullo: {\it B\"acklund transformations for the elliptic Gaudin model and a Clebsch system}, J. Math. Phys., {\bf 52} (2011), 073507.
\end{thebibliography}
\end{document}